\documentclass[preprint,12pt, 3p]{elsarticle}

\usepackage{amsmath}
\usepackage{amssymb}
\usepackage{mathrsfs}
\usepackage{hyperref}
\usepackage{bm}
\usepackage{accents}
\usepackage{graphicx}
\usepackage{wrapfig}
\usepackage[table]{xcolor}
\usepackage{color,soul}
\usepackage{lineno}
\usepackage{graphicx}
\usepackage{subcaption}

\DeclareMathOperator*{\argmin}{arg\,min}

\newdefinition{rmk}{Remark}

\journal{arXiv}

\begin{document}

\begin{frontmatter}

\title{Capturing the fractocohesive length scale through a gradient-enhanced damage model for elastomers}

\author[address1]{S. Mohammad Mousavi}
\author[address2,address3]{Jason Mulderrig}
\author[address4]{Brandon Talamini}
\author[address1,address5]{Nikolaos Bouklas\corref{corr}}
\ead{nb589@cornell.edu}
\address[address1]{Sibley School of Mechanical and Aerospace Engineering, \\ Cornell University, Ithaca, NY 14853, USA}
\address[address2]{Materials and Manufacturing Directorate, Air Force Research Laboratory, Wright-Patterson Air Force Base, Dayton, OH 45433, USA}
\address[address3]{National Research Council (NRC) Research Associateship Programs, The National Academies of Sciences, Engineering, and Medicine, 500 Fifth St., N.W., Washington, DC, 20001, USA}
\address[address4]{Lawrence Livermore National Laboratory, Livermore, CA 94550, USA}
\address[address5]{Pasteur Labs, Brooklyn, NY 11205, USA}



\cortext[corr]{Corresponding authors}


\begin{abstract}
This study aims to unravel the micro-mechanical underpinnings of the emergence of the fractocohesive length scale as a central concept in modern fracture mechanics. A thermodynamically consistent damage and fracture model for elastomers is developed, incorporating elements of polymer chain statistical mechanics. This approach enables the direct incorporation of polymer chain response into a continuum gradient enhanced damage formulation, that in turn allows a physically meaningful description of diffuse chain damage and corresponding fracture events. Through a series of numerical experiments, we simulate crack propagation and extract the fracture energy as an output of the model, while keeping track of the micromechanical signatures of diffuse chain damage that accommodate fracture propagation. Furthermore, we investigate flaw sensitivity and demonstrate that when flaw sizes are smaller than a critical length scale, the material response becomes largely insensitive to notch size. Finally, by combining the fracture toughness and the work to rupture, we identify a fractocohesive length of the material, corresponding to the full width of the damage zone and representing the region where the irreversible dissipation process (i.e., bond scission) is happening. As this region is dictated in the proposed FED model through the introduction of a length scale associated with the non-local nature of the damage and fracture process, the emerging relationship of the two length scales is discussed, effectively connecting the microscopic characteristics of damage to the effective macroscopic response.
\end{abstract} 

\end{frontmatter}

\section{Introduction}\label{Section:Intro}
Elastomers and hydrogels consist of cross-linked networks of flexible polymer chains, a microscopic architecture that enables them to undergo large deformations. Advanced synthesis strategies have also enhanced the toughness response of these materials. These properties have driven their use in both traditional engineering systems and emerging technologies such as soft robotics and biomedical devices \cite{drury2003hydrogels, nonoyama2016double, gent2012engineering, zhalmuratova2020reinforced}. The wide application of these materials has also motivated significant theoretical and computational research to better understand the fundamental mechanisms that govern their behavior and predict their response both under pure mechanical loading and also in a multiphysics setting \cite{bouklas2012swelling, bouklas2015effect,  bouklas2015nonlinear}. Even though elastomers are extensively used in a variety of applications, they are prone to damage and fracture, which can compromise their performance and limit their use in load-bearing applications \cite{mark2003elastomers, sun2012highly, itskov2016rubber, tehrani2017effect, bai2017fatigue}. Bottom-up design for these materials has emerged as a target for advancing synthesis methods, but a fundamental understanding of the mechanics at the chain or network level is still a limiting factor for such efforts. Recent experimental studies have revealed the complex progression from distributed damage to fracture in these materials, connecting phenomena that span length scales several orders of magnitude apart \cite{zhou2021flaw, yang2019polyacrylamide, slootman2020quantifying, lin2020fracture, lin2021fracture, zheng2022fracture, lei2022network, ducrot2014toughening, morelle20213d, sanoja2021mechanical, slootman2022molecular, ju2024role}. However, conventional phenomenological models often fail to capture underlying micromechanical details of the deformation and damage response \cite{xiang2018general, chen2020mechanically}. These limitations highlight the pressing need for advanced damage modeling frameworks that can accurately capture the multiscale processes governing the cascade from damage to fracture in elastomers. This is especially pressing, given the disconnect between traditional experiments that enable acquiring information tied to the macroscopic response, and a new generation of experimental approaches aiming to probe the micromechanical characteristics of damage.

Over the past few decades, a variety of numerical methods have been developed to model damage and fracture.  The phase-field method, in particular, has seen substantial progress in the context of small strains and linear elasticity, with enhancements in algorithmic efficiency, dynamic fracture modeling, and incorporating additional dissipation mechanisms \cite{miehe2010rate, borden2012phase, ambati2015review, yin2020ductile, kumar2018fracture}. Recent work has extended the phase-field framework to soft materials such as elastomers, where large deformations and near-incompressibility must be considered, often requiring Lagrange multipliers that introduce additional numerical difficulties \cite{li2020variational, swamynathan2022phase, feng2023phase, vassilevski1996preconditioning, loghin2004analysis}. While phase-field models provide a robust tool for simulating fracture in soft materials, they do not explicitly track information associated with micromechanical damage processes, which are particularly salient to polymer network mechanics. Gradient-enhanced damage (GED) models offer a promising alternative, as they enable the direct incorporation of physically meaningful damage distributions that can eventually be connected to chain-level mechanics and corresponding continuum-scale fracture simulations \citep{mousavi2024evaluating, mousavi2025chain}. 

The Lake-Thomas (LT) theory \citep{lake1967strength} relates fracture toughness to the energy required to break individual polymer chains. While conceptually insightful, this model often falls short in quantitative predictions, particularly in capturing the complexity of the damage-to-fracture transition often apparent in these materials. This quasibrittle behavior manifests in the statistical nature of processes such as polymerization and network formation, which in turn give rise to imperfections and heterogeneity \cite{flory1953principles}. 
Numerous studies have refined the LT framework, which can be grouped into five thematic advances: (i) molecular-informed upgrades that insert explicit chain mechanics, loop opening, and defect corrections \citep{wang2019quantitative,wang2023contribution,wang2024loop,wang2025loop,lu2025quantify,arora2020fracture,barney2022fracture}; (ii) models that highlight non-local energy dissipation and yield connectivity-based scaling laws \citep{deng2023nonlocal,hartquist2025scaling}; (iii) graph-theoretic and coarse-grained approaches linking fracture to network topology and spatial heterogeneity \citep{arora2021coarse,arora2024effect,yu2025shortest,zhang2024predicting,li2020elongation}; (iv) investigations of mixed weak/strong strands that leverage sacrificial scission and bond-strength percolation for toughening \citep{beech2023reactivity,hartquist2025fracture}; and (v) kinetic perspectives that treat bond rupture as a thermally activated, rate-dependent process \citep{fan2024discontinuous,wang2024fresh,persson2024influence}. In this regard, one should note that fracture in elastomers is inherently a multiscale process, governed by the rupture of individual polymer chains that collectively lead to macroscopic failure. Capturing this transition from bond-level to continuum-scale damage requires models that accurately describe chain and network behavior up to rupture. Chain scission is primarily enthalpic, driven by bond stretching \citep{lake1967strength, wang2019quantitative}. Relevant models, such as the freely-jointed chain (FJC) model \citep{kuhn1942beziehungen}, have been extended to account for bond extensibility \cite{smith1996overstretching, mao2017rupture}. Mao \textit{et al.} \cite{mao2017rupture} incorporated these effects into the Helmholtz free energy to improve predictions of rupture behavior. More recently, Buche \textit{et al.} \cite{buche2021chain, buche2022freely} introduced the $u$FJC model, allowing for arbitrary bond potentials. Due to its complexity, Mulderrig \textit{et al.} \cite{mulderrig2023statistical} proposed a simplified composite version that retains physical accuracy while remaining analytically tractable. Their model also accounts for the statistical nature of thermally activated bond scission and provides an upper bound on chain stretch.

Another critical aspect of elastomer fracture is flaw sensitivity: while small cracks or defects can be catastrophic in brittle materials like glasses, soft elastomers often exhibit insensitivity to small flaws. This flaw-insensitive behavior persists until the flaw exceeds a material-specific threshold, commonly referred to as the fractocohesive length scale \citep{zhou2021flaw, yang2019polyacrylamide, chen2017flaw}\footnote{Rather, a family of length scales, if one is to consider damage under combined loading.}. This length scale, denoted by $\ell_{fc}$, is postulated to be directly related to the size of the damage region that accomodates crack propagation \citep{chen2017flaw, long2021fracture}. When the crack length $c$ is smaller than $\ell_{fc}$, failure becomes dominated by bulk properties rather than defect geometry, leading to flaw-insensitive rupture; conversely, for $c \gg \ell_{fc}$, traditional flaw-sensitive behavior emerges. At the molecular scale, fracture is driven by progressive chain scission, as the applied load stretches and eventually breaks polymer chain backbones and cross-links. This process is loading-rate dependent \citep{mulderrig2023statistical, yang2020multiscale} and unfolds over a spatially extended region surrounding the crack tip, often referred to as the dissipation or failure zone. Recent experiments and theoretical models have shown that chain rupture and energy dissipation occur in a diffuse manner, extending across continuum-relevant scales ranging from tens to hundreds of microns, rather than being confined to molecular-scale damage zones \citep{slootman2020quantifying, long2021fracture, white2025mechanical}. Even though experimental advancements have elucidated a clear picture of the micro-level damage during fracture in elastomers, it is still challenging to quantitatively connect microscopic and macroscopic observations. Specifically, connecting macroscopic predictions of the fractocohesive length scale to the microscopic damage distribution has not yet been achieved, as the fractocohesive length scale encapsulates the multiscale competition of damage and fracture, which is challenging to access in the absence of some model assumptions. These insights motivate the use of a nonlocal continuum framework, such as gradient-enhanced damage (GED) models \citep{peerlings1996gradient}, which can more accurately reflect the diffuse, nonlocal, and rate-dependent evolution of damage in polymer networks \citep{li2020variational, verron2017equal, lamont2021rate}.
In this bottom-up modeling approach, fracture and damage can be studied from prescribing polymer chain behavior and macroscopic loading. This perspective makes GED models an appealing nonlocal framework for coupling statistical mechanics with continuum damage modeling. 

Historically, nonlocal strain-based and damage-based GED models have been widely employed in fracture and damage simulations \citep{peerlings1996gradient, peerlings1998gradient}. However, as noted by de Borst and Verhoosel (2016) \citep{de2016gradient}, these models face limitations in fracture prediction; most notably, the broadening of the damage zone. Significant progress has been made in the past decade, particularly through the work of Lorentz \cite{lorentz2011gradient, lorentz2012modelling, lorentz2017nonlocal} and Talamini \textit{et al.} \cite{talamini2018progressive}, who advanced the implementation of damage-based GED models for fracture. Specifically, Talamini \textit{et al.} \cite{talamini2018progressive} proposed a framework for progressive failure in elastomers based on bond scission mechanics, distinguishing between configurational entropy and internal energy contributions to the free energy. More recently, GED models have gained traction in modeling fracture in soft materials. Lamm \textit{et al.} \cite{lamm2024gradient1} developed a comprehensive framework for crack propagation in polymers under large deformations, incorporating a micromorphic global damage variable to overcome the limitations of local models \cite{forest2009micromorphic}. This framework was further extended to account for thermomechanical coupling and viscoelasticity \cite{lamm2024gradient2}. Despite these advancements, the issue of damage zone broadening persists, as demonstrated by Mousavi \textit{et al.} \cite{mousavi2024evaluating} using a stretch-based GED formulation for elastomers. Some recent studies have attempted to mitigate this issue by introducing a relaxation (or interaction) function designed to diminish nonlocal effects in regions that are fully damaged \cite{poh2017localizing, wosatko2021comparison, wosatko2022survey}, and these efforts showed successful results in a linear setting. Finally,  Mousavi \textit{et al.} \cite{mousavi2025chain} resolved this issue in a strongly-nonlinear finite deformation setting by (i) incorporating a statistical mechanics-based upper bound on the local chain stretch and (ii) introducing a simple relaxation function acting on the nonlocal gradient effects. These innovations effectively eliminated unphysical damage zone broadening and enabled accurate simulation of concurrent damage accumulation and crack propagation in elastomers using a stretch-based GED model. However, a remaining limitation in that work was the use of a simple neo-Hookean free energy density, which restricted the ability of the model to fully incorporate insights from polymer chain statistical mechanics.

In this work, we aim to address this concern by utilizing the Helmholtz free energy density derived from a monodisperse elastomer network in the GED model that naturally incorporates a non-local length scale encapsulating network heterogeneity. Moreover, a key objective of this study, starting from assumptions of response and damage at the chain-level, is to determine the predicted fractocohesive length of the material through upscaling via the GED model and corresponding simulation, and to interpret its correspondence to the micro-level damage distribution. This study builds on the theoretical framework and model introduced by Mousavi \textit{et al.} (2025) \citep{mousavi2025chain}. Whereas the previous work primarily centered on developing and validating the theory and numerical implementation in a general setting of finite deformation fro quasibrittle materials, the focus here is on uncovering the fundamental physics that underpin it. The structure of the paper is as follows: Section \ref{Section:formulation} presents the statistical mechanics-based free energy formulation for elastomers and its integration into a thermodynamically consistent GED model. Section \ref{Section:NumericalImplementation} outlines the numerical implementation using the open-source finite element library FEniCS \cite{alnaes2015fenics}. In Section \ref{Section:results}, we first simulate crack propagation using the proposed model and compare the results with those from a phase-field approach. We then investigate flaw-insensitive behavior and extract the material’s fractocohesive length. Finally, we provide an analytical estimate of the fracture toughness based on a modified LT theory and compare it with the numerical predictions. The full implementation is openly accessible on GitHub\footnote{\url{https://github.com/MMousavi98/Statistical_GED}}.

\section{Formulation}\label{Section:formulation}
\begin{figure}[h!]
    \centering
    \includegraphics[width=0.9\linewidth]{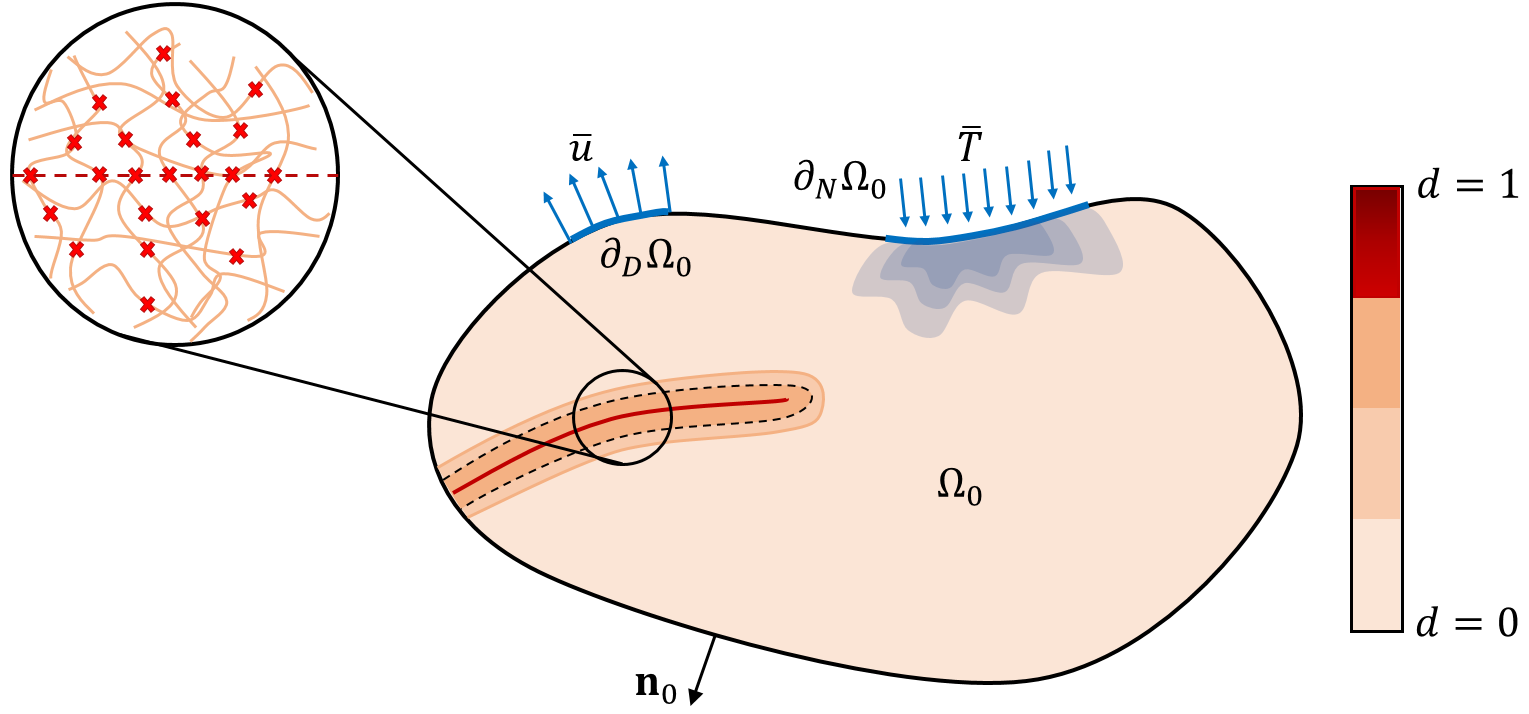}
        \caption{Schematic of a solid continuum that has undergone diffuse damage and fracture.}
    \label{fig:potato}
\end{figure}

Throughout all formulations and modeling procedures, the reference configuration ${\Omega}_0 \subset \mathbb{R}^3$ is assumed to be an open, bounded, and connected domain with a sufficiently smooth boundary, denoted by $\partial \Omega_0$. The solid body is illustrated in Fig. \ref{fig:potato}, which also depicts the imposed boundary conditions, a sharp crack discontinuity, and a surrounding region of diffuse damage. As shown, Dirichlet and Neumann boundary conditions are prescribed on $\partial_D\Omega_0$ and $\partial_N\Omega_0$, respectively, where $\partial_D\Omega_0 \subset \Omega_0$ and $\partial_N\Omega_0 \subset \Omega_0$.

\subsection{Kinematics}\label{SubSection:Kinematics}

When the body deforms, each material point transitions from its original position $\bm{X}$ in the reference configuration $\Omega_0$ to a new position $\bm{x}$ in the deformed configuration $\Omega$. This motion is described by the displacement vector $\bm{u}$, defined as $\bm{u} = \bm{x} - \bm{X}$, which represents the change in position of a particle from its referential to current state. To fully characterize the deformation, it is essential to examine how infinitesimal line elements transform, which is captured by the relation:
\begin{equation}\label{bulk-element}
d\bm{x} = \mathbf{F}(\bm{X},t)\, d\bm{X},
\end{equation}
where $\mathbf{F}$ is the deformation gradient, given by
\begin{equation}\label{deformation-gradient}
\mathbf{F}(\bm{X},t) = \frac{\partial \bm{x}}{\partial \bm{X}}.
\end{equation}
The Jacobian determinant $J(\bm{X}, t) = dv/dV = \det \mathbf{F}(\bm{X}, t)$ quantifies the local volume change between the deformed and reference configurations and must remain positive for physically admissible deformations.

In addition, the right Cauchy-Green deformation tensor $\mathbf{C}$ is introduced as
\begin{equation}
\mathbf{C} = \mathbf{F}^T \mathbf{F},
\end{equation}
whose principal invariants are defined as follows \cite{holzapfel2002nonlinear}:
\begin{equation} \label{invariants}
\begin{split}
    I_1 = \mathrm{tr}(\mathbf{C}), \\
    I_2 = \frac{1}{2}\left[(\mathrm{tr}(\mathbf{C}))^2 - \mathrm{tr}(\mathbf{C}^2)\right], \\
    I_3 = \det(\mathbf{C}).
\end{split}
\end{equation}

\subsection{Free energy of elastomeric chain networks}\label{SubSection:chainenergetics}
In this section, we follow the approach of \cite{li2020variational} and  \cite{talamini2018progressive}. To overcome the limitations of the classical Gaussian chain model, we adopt a non-Gaussian formulation \cite{kuhn1942beziehungen,james1943theory,treloar1975physics} that captures the finite extensibility of polymer chains as the end-to-end distance $r$ approaches the contour length $Nb$, where $N$ is the number of Kuhn segments and $b$ is the segment length. The Helmholtz free energy of an individual chain composed of $N$ freely jointed segments of length $b$ is given by \citep{flory1953principles,kuhn1942beziehungen,james1943theory,treloar1975physics}:
\begin{equation}
\label{free_energy_single_chain}
\psi(r) = k_b\theta N\left(\frac{r}{Nb}\beta+\ln\frac{\beta}{\sinh\beta}\right), \quad \beta = \mathscr{L}^{-1}\left(\frac{r}{Nb}\right),
\end{equation}
where $k_b$ is the Boltzmann constant, $\theta$ is the absolute temperature, and $\mathscr{L}^{-1}$ is the inverse Langevin function associated with the Langevin function $\mathscr{L}(x) = \coth(x) - 1/x$. An additive constant in the free energy is omitted since it does not influence the material response. The corresponding non-Gaussian tensile force along the direction of $r$ is expressed as:
\begin{equation}
\label{single_chain_force}
f = \frac{k_b\theta}{b} \, \mathscr{L}^{-1}\left(\frac{r}{Nb}\right).
\end{equation}

Polymer network models grounded in the statistical mechanics of individual long-chain molecules \cite{treloar1979non, arruda1993three, wu1993improved, boyce2000constitutive} have demonstrated strong agreement with experimental observations across various large-strain deformation modes, including uniaxial tension, equibiaxial stretching, and pure shear. These models rely on the non-Gaussian chain formulation, where the chain stretch approaches its contour length ($r \rightarrow Nb$), leading to a divergence in both the free energy \eqref{free_energy_single_chain} and the chain force \eqref{single_chain_force}. As a result, the validity of the non-Gaussian chain model is inherently limited to the regime $r < Nb$, and as a consequence cannot accurately capture chain scission. 

To capture the rupture of elastomer networks via scission of backbone bonds, Mao \textit{et al.} \cite{mao2017rupture} extended the classical chain model by incorporating concepts originally proposed by Smith \textit{et al.} \cite{smith1996overstretching} for modeling the overstretching behavior of DNA molecules. Their approach relaxes the traditional rigidity assumption of Kuhn segments by accounting for both segmental deformation and alignment under tensile loading. In this modified model, the free energy of a single chain is augmented to include the energetic contribution from stretching the Kuhn segments, and is expressed as \citep{mao2017rupture}:
\begin{equation}
\label{free_energy_single_chain_extend0}
\psi_c\left(r,\lambda_\mathrm{b}\right) = U(\lambda_\mathrm{b}) + k_b\theta N\left(\frac{r}{N \lambda_\mathrm{b}b}\beta+\ln\frac{\beta}{\sinh\beta}\right), \quad \beta = \mathscr{L}^{-1}\left(\frac{r}{N \lambda_\mathrm{b}b}\right),
\end{equation}
where $\lambda_\mathrm{b}$ is the Kuhn segment stretch ratio, and $U(\lambda_\mathrm{b})$ denotes the internal energy stored in the stretched Kuhn segments.

At the single chain level, chain stretch, $\lambda_\mathrm{ch}$, is defined as the ratio between the deformed and undeformed chain lengths, $\lambda_\mathrm{ch} = r / r_0$, where $r_0$ is the equilibrium chain length. Here, we take $r_0$ to be equal to the expected distance of a random walk of $N$ steps, each of length $b$; $r_0 = \sqrt{N}b$. 
Accordingly, the chain free energy in Eq. \eqref{free_energy_single_chain_extend0} can be reformulated as a function of $\lambda_\mathrm{ch}$ and $\lambda_\mathrm{b}$ as:
\begin{equation}
\label{free_energy_single_chain_extend}
\psi_c\left(\lambda_\mathrm{ch},\lambda_\mathrm{b}\right) = U(\lambda_\mathrm{b}) + k_b\theta N\left(\frac{\lambda_\mathrm{ch}\lambda_\mathrm{b}^{-1}}{\sqrt{N}}\beta+\ln\frac{\beta}{\sinh\beta}\right), \quad \beta = \mathscr{L}^{-1}\left(\frac{\lambda_\mathrm{ch}\lambda_\mathrm{b}^{-1}}{\sqrt{N}}\right).
\end{equation}

A comparison between the extensible free energy functional \eqref{free_energy_single_chain_extend} and the classical form \eqref{free_energy_single_chain} reveals that the modified stretch term $\lambda_\mathrm{ch}\lambda_\mathrm{b}^{-1}$ can be interpreted as the effective chain stretch arising solely from the reorientation and alignment of Kuhn segments, excluding contributions from bond stretching \cite{mao2017rupture, talamini2018progressive}. In the following, we utilize a simple quadratic form for the internal energy of the chain, as proposed in \cite{talamini2018progressive} and \cite{mao2018theory}:
\begin{equation}
\label{single_chain_internal_energy}
U(\lambda_\mathrm{b}) = \frac{1}{2} N E_\mathrm{b}\left( \lambda_\mathrm{b} - 1\right)^2,
\end{equation}
where $E_\mathrm{b}$ denotes the stiffness of the chemical bonds in the chain backbone. 

For a prescribed chain stretch $\lambda_\mathrm{ch} = \bar{\lambda}_\mathrm{ch}$, and while neglecting thermal fluctuations\footnote{The reader is directed to \cite{mulderrig2023statistical} for a more extensive discussion}, the corresponding Kuhn segment stretch $\lambda_\mathrm{b}$ results from a balance between the entropic and energetic contributions to the total free energy \eqref{free_energy_single_chain_extend}. Specifically, $\lambda_\mathrm{b}$ is obtained by minimizing the free energy with respect to segment stretch:
\begin{equation}
\lambda_\mathrm{b} = \argmin_{\lambda_\mathrm{b} \ge 1} \tilde{\psi}_c(\lambda_\mathrm{b}), \quad \text{where} \quad \tilde{\psi}_c(\lambda_\mathrm{b}) := \psi_c(\lambda_\mathrm{ch}, \lambda_\mathrm{b}) \quad \text{with} \quad \lambda_\mathrm{ch} = \bar{\lambda}_\mathrm{ch}.
\end{equation}
More details about the solution of this minimization problem are fleshed out in \cite{li2020variational}.


Starting from the single-chain free energy \eqref{free_energy_single_chain_extend} and considering the chain density $n$ (the number of chains per unit reference volume), the Helmholtz free energy density of the elastomer network   can be expressed as:
\begin{equation}
\label{network_free_energy}
\psi\left(\bm{u},\lambda_\mathrm{b}\right) = \frac{1}{2} N E \left(\lambda_\mathrm{b} - 1\right)^2 + N\mu\left(\frac{\lambda_\mathrm{ch}\lambda_\mathrm{b}^{-1}}{\sqrt{N}} \beta + \ln\frac{\beta}{\sinh\beta}\right), \quad \beta = \mathscr{L}^{-1}\left(\frac{\lambda_\mathrm{ch}\lambda_\mathrm{b}^{-1}}{\sqrt{N}}\right),
\end{equation}
where $\mu = nk_b\theta$ is the shear modulus of the network, and $E = nE_\mathrm{b}$ represents the stiffness associated with bond stretching across the polymer network. Additionally, to connect chain stretch to the continuum deformation via a macro-to-micro mapping, the 8-chain model \cite{Arruda-Boyce1993} suggests that the chain stretch at a material point is related to the first invariant of the right Cauchy-Green deformation tensor as:
\begin{equation}\label{Lambda_chain}
\lambda_\mathrm{ch} = \sqrt{\frac{I_1}{3}}.
\end{equation}
This model (Eq. \eqref{network_free_energy}) will be used in subsection \ref{subsection:model_specialization}.

\subsection{Thermodynamic considerations}

Following the developments from Section \ref{SubSection:chainenergetics} and adapting to the procedure presented in \cite{mousavi2025chain} for the development of a GED model -- where here we assign nonlocal (gradient) effects to a nonlocal segment stretch $\Bar{\lambda}_b$ -- then the internal mechanical power $P_{int}$ for a nearly incompressible polymer network in the reference configuration $\Omega_0$ is expressed as
\begin{equation}\label{eq::rateIntPower}
    P_{int} = \int_{\Omega_0}\left[\mathbf{P}\colon\nabla\dot{\bm{u}} + f_{p}\dot{p} + f_{\Bar{\lambda}_\mathrm{b}}\dot{\Bar{\lambda}}_\mathrm{b} + \boldsymbol{\xi}_{\Bar{\lambda}_\mathrm{b}}\cdot\nabla\dot{\Bar{\lambda}}_\mathrm{b} + f_{d}\dot{d}\right]dV,
\end{equation}
where displacement $\mathbf{u}$, pressure $p$, nonlocal segment stretch $\Bar{\lambda}_b$, and damage $d$ are considered the state variables of the problem. The nonlocal segment stretch $\Bar{\lambda}_b$ is considered here, along with the nonlocal length scale $\ell$, to provide augmented information regarding the microscopic deformation of the continuum, taking into account network heterogeneity and imperfections that influence load-sharing. Here, $\mathbf{P}$ is the first Piola-Kirchhoff stress tensor, and $f_{p}$, $f_{\Bar{\lambda}_\mathrm{b}}$, $\boldsymbol{\xi}_{\Bar{\lambda}_\mathrm{b}}$, and $f_{d}$ represent power conjugates to the internal state variables and corresponding gradients. By neglecting body forces, the external mechanical power $P_{ext}$ can be expressed as follows,
\begin{equation}
    P_{ext} = \dot{W}_{ext} = \int_{\partial_N\Omega_0}{\bm{T}}\cdot\dot{\bm{u}} dA + \int_{\partial_N\Omega_0}\hat{\iota}\dot{\Bar{\lambda}}_\mathrm{b} dA,
\end{equation}
where $\bm{T}$ is the mechanical surface traction and $\hat{\iota}$ defines micromorphic reactions. By satisfying the principle of virtual power (first law of thermodynamics in variational form, $\delta{P}_{int} = \delta{P}_{ext}$) and then applying the divergence theorem, the following governing equations and boundary conditions are obtained:
\begin{itemize}
    \item Mechanical equilibrium equation and boundary conditions: \begin{equation} \label{mechanical_equilibrium_governing}
    \begin{split}
        \nabla\cdot\mathbf{P}= \mathbf{0} \quad \text{in} \quad \Omega_0,\\
        \bm{u}= \Bar{\bm{u}} \quad \text{on} \quad \partial_{D}{\Omega}_0,\\
        \mathbf{P}\cdot\bm{n}_0 = \bm{T} \quad \text{on} \quad \partial_N\Omega_0.
        \end{split}
    \end{equation}
    \item {Power conjugate of pressure:}
    \label{pressure_governing}
    \begin{equation}
        f_{p} = 0.
    \end{equation}
    \item Micro-force equilibrium equation and boundary conditions: \begin{equation} \label{chain_governing}
    \begin{split}
        \nabla\cdot\boldsymbol{\xi}_{\Bar{\lambda}_\mathrm{b}}-f_{\Bar{\lambda}_\mathrm{b}} = 0 \quad \text{in} \quad \Omega_0,\\
        \Bar{\lambda}_\mathrm{b}= \Bar{\Bar{\lambda}}_\mathrm{b} \quad \text{on} \quad \partial_{D}{\Omega}_0,\\
        \qquad\boldsymbol{\xi}_{\Bar{\lambda}_\mathrm{b}}\cdot\bm{n}_0 = \hat{\iota} \quad \text{on} \quad \partial_{N}{\Omega}_0.
        \end{split}
    \end{equation}
    \item {Power conjugate of damage:}
    \begin{equation} \label{damage_governing1}
        f_{d} = 0.
    \end{equation}
\end{itemize}

The Helmholtz free energy is expressed as a function of both external and internal state variables, given by $\Psi(\mathbf{F}, p, \Bar{\lambda}_\mathrm{b}, \nabla\Bar{\lambda}_\mathrm{b}, d)$. Based on this form, the material time derivative of the free energy can be obtained using the chain rule as
\begin{equation}\label{eq::rateFreeEnergy}
    \frac{\textnormal{d}\Psi}{\textnormal{d} t} = \frac{\partial\Psi}{\partial\mathbf{F}}\colon\dot{\mathbf{F}} + \frac{\partial \Psi}{\partial p}\dot{p} + \frac{\partial \Psi}{\partial \Bar{\lambda}_\mathrm{b}}\dot{\Bar{\lambda}}_\mathrm{b} + \frac{\partial \Psi}{\partial \nabla\Bar{\lambda}_\mathrm{b}}\cdot\nabla\dot{\Bar{\lambda}}_\mathrm{b} + \frac{\partial \Psi}{\partial d}\dot{d}.
\end{equation}

To ensure consistency with the second law of thermodynamics, we apply the Coleman-Noll procedure and combine the above with the internal power expression in Eq. \eqref{eq::rateIntPower}, leading to the following expression for the rate of energy dissipation:
\begin{equation}\label{rateDis}
   \mathcal{D} = \left(\mathbf{P} - \frac{\partial\Psi}{\partial\mathbf{F}}\right)\colon\dot{\mathbf{F}} + \left(f_p - \frac{\partial \Psi}{\partial p}\right)\dot{p} + \left(f_{\Bar{\lambda}_\mathrm{b}} - \frac{\partial \Psi}{\partial \Bar{\lambda}_\mathrm{b}}\right)\dot{\Bar{\lambda}}_\mathrm{b} + \left(\boldsymbol{\xi}_{\Bar{\lambda}_\mathrm{b}} - \frac{\partial \Psi}{\partial \nabla\Bar{\lambda}_\mathrm{b}}\right)\cdot\nabla\dot{\Bar{\lambda}}_\mathrm{b} + \left(f_d - \frac{\partial \Psi}{\partial d}\right)\dot{d} \geq 0.
\end{equation}

To ensure that the inequality holds for arbitrary rates of the state variables, the following constitutive relations are obtained for the first four terms:
\begin{equation}\label{eq::constitutive}
    \mathbf{P} = \frac{\partial\Psi}{\partial\mathbf{F}}, \quad
    f_p = \frac{\partial \Psi}{\partial p}, \quad
    f_{\Bar{\lambda}_\mathrm{b}} = \frac{\partial \Psi}{\partial \Bar{\lambda}_\mathrm{b}}, \quad
    \boldsymbol{\xi}_{\Bar{\lambda}_\mathrm{b}} = \frac{\partial \Psi}{\partial \nabla\Bar{\lambda}_\mathrm{b}} \quad \text{in } \Omega_0.
\end{equation}

For the fifth term in Eq. \eqref{rateDis}, we assume $f_d = 0$, consistent with Eq. \eqref{damage_governing1}. Inserting this into Eq. \eqref{rateDis}, the dissipation associated with damage becomes
\begin{equation}\label{eq:reduced-cp-dissipation-inequality-version-1}
   \mathcal{D} = -\frac{\partial \Psi}{\partial d}\dot{d} \geq 0.
\end{equation}


Since damage is considered an irreversible process, the damage variable $d$ is assumed to be monotonically increasing over time, i.e., $\dot{d} \geq 0$. As a result, in order to satisfy the dissipation inequality, it must hold that
\begin{equation} \label{eq:reduced-cp-dissipation-inequality-version-2}
\frac{\partial \Psi}{\partial d} \leq 0 \quad \text{in } \Omega_0,
\end{equation}
which places a constraint on the constitutive form of the free energy function.

\subsection{Model specialization}\label{subsection:model_specialization}
Again in line with \cite{mousavi2025chain}, to specialize the Helmholtz free energy for near-incompressible elastomeric materials, we decompose the total free energy into local and nonlocal components as $\Psi = \Psi^{\text{loc}} + \Psi^{\text{nloc}}$. The local contribution captures the elastic response degraded by damage. To enforce near-incompressibility, we adopt a perturbed Lagrangian formulation \cite{li2020variational,wriggers2006computational,ang2022stabilized}, introducing a pressure-like field $p$ and a penalty parameter $\kappa$ which regulates compressibility. The resulting local free energy density is given by:
\begin{equation}\label{psi_loc}
    \Psi^{\text{loc}}(\bm{u},\Bar{\lambda}_\mathrm{b},p,d) = a(d)\psi(\bm{u},\Bar{\lambda}_\mathrm{b}) - b(d)p(J - 1) - \frac{p^2}{2\kappa},
\end{equation}
where $\psi(\bm{u},\Bar{\lambda}_\mathrm{b})$ is the network-level free energy defined in Eq. \eqref{network_free_energy}, and $\Bar{\lambda}_\mathrm{b}$ denotes the nonlocal Kuhn segment stretch. The degradation functions are defined as:
\begin{equation}\label{ab}
\begin{split}
a(d) &= (1-k_{\ell})(1-d)^2 + k_{\ell}, \\
b(d) &= (1-k_{\ell})(1-d)^3 + k_{\ell},
\end{split}
\end{equation}
and $k_{\ell}$ is a small numerical parameter for regularization. The higher-order polynomial in $b(d)$ ensures that the effective resistance to bulk deformation decreases more rapidly than shear resistance, allowing crack-induced volume expansion while maintaining incompressibility in the undamaged regions \citep{ang2022stabilized}.

\begin{figure}[h!]
    \centering
    \includegraphics[width=0.6\linewidth]{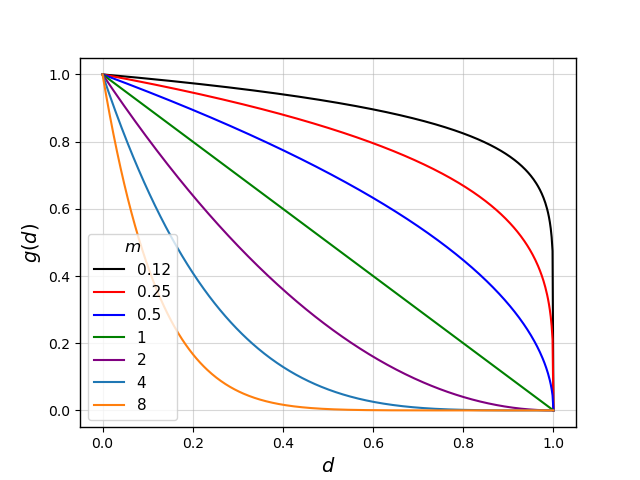}
    \caption{The effect of different relaxation function parameters $m$ on the relaxation function $g(d)$.}
    \label{fig:g_plot}
\end{figure}

The nonlocal energy decomposed as $\Psi^{\text{nloc}} = \Psi_{micmac}^{nloc} + \Psi_{grd}^{nloc}$ and captures network-level effects. The micro-macro coupling term penalizes deviations between the local Kuhn segment stretch $\lambda_\mathrm{b}$ and the nonlocal Kuhn segment stretch $\Bar{\lambda}_\mathrm{b}$:
\begin{equation} \label{psi_nloc_micmac}
    \Psi_{micmac}^{nloc}(\lambda_\mathrm{b}, \Bar{\lambda}_\mathrm{b}) = \frac{h_{nl}}{2} \left[\lambda_\mathrm{b} - \Bar{\lambda}_\mathrm{b}\right]^2,
\end{equation}
where $h_{nl}>0$ controls the intensity of nonlocal interactions. The gradient regularization term accounts for spatial variations in the nonlocal stretch field:
\begin{equation} \label{psi_nloc_grd}
    \Psi_{grd}^{nloc}(\nabla\Bar{\lambda}_\mathrm{b},d) = \frac{h_{nl}g(d)\ell^2}{2} \nabla\Bar{\lambda}_\mathrm{b} \cdot \nabla\Bar{\lambda}_\mathrm{b},
\end{equation}
where $g(d)$ is a relaxation function and $\ell$ corresponds to the physically motivated nonlocal length scale of the material that manifests due to network level heterogeneities and imperfections. \footnote{It is noted that in the current formulation, the parameter $h_{nl}$ enters both Eqs. \ref{psi_nloc_micmac} and \ref{psi_nloc_grd}. We have not explored scenarios where energy contributions from these two equations are weighed differently.}. Inspired by previous works \cite{poh2017localizing, wosatko2021comparison, wosatko2022survey}, this function is defined as $g(d) = (1 - d)^m$, with $m > 0$ controlling the rate of decay. It progressively reduces nonlocal interactions as damage accumulates, ensuring that fully deteriorated regions no longer influence the surrounding material. As illustrated in Fig. \ref{fig:g_plot}, smaller values of $m$ lead to a sharper drop in interaction strength at high damage levels, helping to avoid nonphysical residual stresses in the damage zone. Further details on the derivation and calibration of this framework can be found in our previous work \cite{mousavi2025chain}.

\subsubsection{Constitutive relations}
Following Eqs. \eqref{psi_loc}, \eqref{psi_nloc_micmac}, and \eqref{psi_nloc_grd}, the Helmholtz free energy is given by
\begin{align} \label{eq::Helmholtz}
    \Psi(\bm{u},p, \Bar{\lambda}_\mathrm{b}, \nabla\Bar{\lambda}_\mathrm{b}, d) = & a(d)\psi(\bm{u},\Bar{\lambda}_\mathrm{b}) -b(d)p\left(J -1\right) -\frac{p^2}{2\kappa} \nonumber \\
    & + \frac{h_{nl}}{2}\left[\lambda_\mathrm{b} - \Bar{\lambda}_\mathrm{b}\right]^2 + \frac{h_{nl}g(d)\ell^2}{2} \nabla\Bar{\lambda}_\mathrm{b}\cdot\nabla\Bar{\lambda}_\mathrm{b}.
\end{align}
and fulfills the condition imposed by Eq. \eqref{eq:reduced-cp-dissipation-inequality-version-2}. By combining Eqs. \eqref{eq::constitutive} and \eqref{eq::Helmholtz}, we arrive at the following constitutive expressions:

\begin{equation}\label{constitutives}
\begin{split}
    \mathbf{P} = a(d)\frac{\partial\psi}{\partial\mathbf{F}} - b(d) p J \mathbf{F}^{-T} \quad \text{in} \quad \Omega_0, \\
    f_p=-b(d)(J-1)-\frac{p}{k} \quad \text{in} \quad \Omega_0, \\
    \boldsymbol{\xi}_{\Bar{\lambda}_\mathrm{b}} = h_{nl}g(d)\ell^2\nabla\Bar{\lambda}_\mathrm{b}\quad \text{in} \quad \Omega_0, \\
    f_{\Bar{\lambda}_\mathrm{b}} =  -h_{nl}\left[\lambda_\mathrm{b} - \Bar{\lambda}_\mathrm{b}\right] \quad \text{in} \quad \Omega_0.
\end{split}
\end{equation}

\subsubsection{Damage function}
In strain- or stretch-based GED models, the damage function must be expressed in terms of a nonlocal variable. For rate-dependent formulations, the rate of this nonlocal quantity may also enter the damage evolution law. In our thermodynamically consistent framework, we define a constitutive expression for the thermodynamic conjugate force associated with damage as
\begin{equation}\label{damage_conjugateforce}
f_d(d,\Bar{\lambda}_\mathrm{b}) = d - \frac{1}{1+\textnormal{exp}(-c(\Bar{\lambda}_\mathrm{b}-\lambda_\mathrm{cr}))},
\end{equation}
where $\lambda_\mathrm{cr}$ denotes the critical (Kuhn segment) stretch corresponding to the midpoint of the damage transition ($d = 0.5$), and $c > 0$ controls the steepness of the sigmoid function. A larger value of $c$ results in a sharper transition from the undamaged ($d \approx 0$) to the fully damaged state ($d \approx 1$), while a smaller $c$ leads to a more gradual evolution of damage. This is in line with previous developments, such as \cite{mulderrig2021affine}, which aims to encapsulate the effect of network imperfections and heterogeneity in the proposed damage criterion. 

The damage evolution is also influenced by a history variable $\mathcal{H}$, which tracks the maximum value attained by the nonlocal stretch over time:
\begin{equation}\label{history}
\mathcal{H}(\Bar{\lambda}_\mathrm{b};\mathbf{X},t)=\mathrm{max}_{s\in[0,t]}\Bar{\lambda}_\mathrm{b}(\mathbf{X},s)\,.
\end{equation}

By applying the balance law given in Eq. \eqref{damage_governing1}, Eq. \eqref{damage_conjugateforce} can be rearranged to recover the commonly used form of the damage function in GED models:
\begin{equation}\label{damage_function}
d = \frac{1}{1+\textnormal{exp}(-c(\Bar{\lambda}_\mathrm{b}-\lambda_\mathrm{cr}))}.
\end{equation}
This expression corresponds to the strong-form solution of the damage balance equation and inherently satisfies the irreversibility condition ($\dot{d} \geq 0$).

Various other damage functions have been proposed in the literature \cite{mulderrig2023statistical, peerlings1996gradient, verhoosel2011isogeometric, sarkar2019comparative}. Compared to the damage function employed in our previous work \cite{mousavi2025chain}, the current sigmoid formulation offers two key advantages. First, it is simpler, involving only a single parameter $c$ rather than the two parameters $c$ and $\gamma$ used previously. Second, its smoothness (consistent with insight from statistical mechanics \citep{mulderrig2023statistical}) enhances numerical stability during simulations.

\subsection{Strong and weak forms} \label{section:strong_weak_forms}
The micro-force balance equation and its corresponding Neumann boundary condition, originally given in Eq. \eqref{chain_governing}, can be reformulated as:
\begin{equation} \label{chain_strong_boundary_1}
\begin{split}
    h_{nl}\left[\lambda_\mathrm{b} - \Bar{\lambda}_\mathrm{b} + \nabla\cdot\left(g(d)\ell^2 \nabla\Bar{\lambda}_\mathrm{b}\right)\right] = 0\quad \text{in} \quad \Omega_0, \\
    h_{nl}g(d)\ell^2\nabla\Bar{\lambda}_\mathrm{b}\cdot\bm{n}_0 = \hat{\iota} \quad \text{on} \quad \partial_{N}{\Omega}_0.
\end{split}
\end{equation}
Since $h_{nl}$ is strictly positive, it can be factored out from both the bulk and boundary equations in Eq. \eqref{chain_strong_boundary_1} \cite{peerlings2004thermodynamically}. Furthermore, in line with the physical interpretation and following the argument by Sarkar \textit{et al.} (2020) \cite{sarkar2020thermo}, the boundary micro-traction $\hat{\iota}$ can be considered negligible. Applying these simplifications and incorporating the constitutive laws into the governing equations in Eqs. \eqref{mechanical_equilibrium_governing} and \eqref{chain_governing}, we obtain the following set of strong form equations:
\begin{equation}
\begin{split}\label{strong_forms}
    \nabla\cdot\left[a(d)\frac{\partial\psi}{\partial\mathbf{F}} - b(d) p J \mathbf{F}^{-T}\right] = \mathbf{0} \quad \text{in} \quad \Omega_0, \\
    - b(d)(J - 1) - \frac{p}{\kappa} = 0 \quad \text{in} \quad \Omega_0, \\
    \Bar{\lambda}_\mathrm{b} - \lambda_\mathrm{b} - \nabla\cdot\left(g(d)\ell^2 \nabla\Bar{\lambda}_\mathrm{b}\right) = 0 \quad \text{in} \quad \Omega_0.
\end{split}
\end{equation}
with the corresponding boundary conditions
\begin{equation}\label{BC}
\begin{split}
\bm{u} = \Bar{\bm{u}} \quad \text{on} \quad \partial_{D}{\Omega}_0, \\
\mathbf{P}\cdot\bm{n}_0 = \bm{T} \quad \text{on} \quad \partial_N\Omega_0, \\
\Bar{\lambda}_\mathrm{b}= \Bar{\Bar{\lambda}}_\mathrm{b} \quad \text{on} \quad \partial_{D}{\Omega}_0, \\
\nabla\Bar{\lambda}_\mathrm{b}\cdot\bm{n}_0 = 0 \quad \text{on} \quad \partial_{N}{\Omega}_0.
\end{split}
\end{equation}

During monotonic loading, as damage evolves, the local Kuhn segment stretch $\lambda_\mathrm{b}$ continues to increase within the damaged region, even after a crack has initiated and begun to propagate. Consequently, the driving force in Eq. $\eqref{strong_forms}_3$ remains active beyond fracture, leading to the ongoing, non-physical growth of the nonlocal Kuhn segment stretch $\Bar{\lambda}_\mathrm{b}$ and resulting in the broadening of the damage zone \cite{mousavi2024evaluating}. To address this issue, the relaxation function $g(d)$ has been proposed; however, previous studies have shown no definitive evidence that it fully resolves the damage zone broadening problem \cite{wosatko2021comparison, wosatko2022survey}. More recently, Mousavi \textit{et al.} (2025) demonstrated that the relaxation function alone is insufficient, and additional considerations are necessary to effectively mitigate this behavior.

Polymer statistical mechanics offers insight to this situation by framing chain scission as a thermally activated process. As discussed in previous work of some of the authors \citep{mulderrig2023statistical} and in the work of Wang \textit{et al.} (2019) \cite{wang2019quantitative}, the activation energy for scission depends on the applied chain stretch. Above a certain stretch value, the energy barrier for chain scission effectively vanishes, indicating a limit beyond which no chains can survive.\footnote{We remind that the formulation developed in this work only considers the limit of no thermal fluctuations. Thus, progressive damage is due to the proposed damage law, as motivated by network level flaws.} We denote this threshold as $\lambda_\mathrm{b}^{\text{max}}$, which serves as a physically meaningful upper bound for the Kuhn segment stretch influencing the evolution of $\Bar{\lambda}_\mathrm{b}$.\footnote{Note that $\lambda_\mathrm{b}^{\text{max}}$ is distinct from $\lambda_\mathrm{cr}$; the former represents the maximum admissible Kuhn segment stretch that governs the nonlocal field, whereas the latter marks the stretch at which $d=0.5$ (essentially corresponding to the midpoint shift of the sigmoid function).}

To incorporate this physical constraint into the model, we modify the expression in Eq. \eqref{psi_nloc_micmac} by replacing $\lambda_\mathrm{b}$ with $\lambda_\mathrm{b}^{\text{max}} - \ll\lambda_\mathrm{b}^{\text{max}} - \lambda_\mathrm{b}\gg$. This modification adjusts the expression for $f_{\Bar{\lambda}}$ in Eq. $\eqref{constitutives}_4$, and results in the following reformulated strong form:
\begin{equation}\label{GED_strong2}
\Bar{\lambda}_\mathrm{b} - \lambda_\mathrm{b}^{\text{max}} + \ll\lambda_\mathrm{b}^{\text{max}} - \lambda_\mathrm{b}\gg - \nabla\cdot\left(g(d)\ell^2 \nabla\Bar{\lambda}_\mathrm{b}\right) = 0 \quad \text{in } \Omega_0.
\end{equation}
Note the application of Macaulay brackets, which are defined by the following operation:
\begin{equation}
\ll x \gg = 
\begin{cases}
x, & \text{if } x > 0 \\
0, & \text{if } x \leq 0
\end{cases}
\label{eq:macaulay_brackets}
\end{equation}
This modification imposes an upper bound on $\lambda_\mathrm{b}$ in Eq. \eqref{GED_strong2}, preventing it from exceeding $\lambda_\mathrm{b}^{\text{max}}$ in an unphysical manner. The approach is consistent with the statistical mechanics framework for polymer chains, where damage is stretch-dependent, similar to continuum damage mechanics \cite{mulderrig2021affine}. In statistical models that consider ensembles of chains, a critical stretch naturally arises (serving as a ``cut-off'') beyond which further deformation has no effect on damage progression, as all chains are presumed to have failed. This idea carries over to the continuum formulation by introducing a similar cut-off for the damaged driving force (i.e., the local Kuhn segment stretch), thereby ensuring physical consistency in the model.

To formulate the weak form of the governing equations, we introduce the trial functions $(\bm{u}, p, \lambda) \in (\mathbb{U}, \mathbb{P}, \mathbb{L})$, where the associated function spaces are defined as:
\begin{equation}\label{trial_spaces}
\begin{split}
\mathbb{U} = \{\bm{u} \in H^1(\Omega_0)\;|\; \bm{u} = \Bar{\bm{u}} \text{ on } \partial_D\Omega_0\}, \\
\mathbb{P} = \{p \in L^2(\Omega_0)\}, \\
\mathbb{L} = \{\lambda \in H^1(\Omega_0)\;|\; \lambda = \Bar{\Bar{\lambda}} \text{ on } \partial_D\Omega_0\},
\end{split}
\end{equation}
where $H^1$ denotes the first-order Sobolev space and $L^2$ represents the space of square-integrable functions. Correspondingly, we define the test functions $(\bm{v}, q, \beta) \in (\mathbb{V}, \mathbb{Q}, \mathbb{B})$ within the following spaces:
\begin{equation}\label{test_spaces}
\begin{split}
\mathbb{V} = \{\bm{v} \in H^1_0(\Omega_0)\;|\; \bm{v} = \bm{0} \text{ on } \partial_D\Omega_0\}, \\
\mathbb{Q} = \{q \in L^2(\Omega_0)\}, \\
\mathbb{B} = \{\beta \in H^1_0(\Omega_0)\;|\; \beta = 0 \text{ on } \partial_D\Omega_0\}.
\end{split}
\end{equation}
By applying the standard Galerkin procedure, we derive the following weak form:
\begin{equation}\label{weak_form_GED}
\begin{split}
    \int_{\Omega_0} \left[a(d)\frac{\partial\psi}{\partial\mathbf{F}} - b(d)\,p\,J\,\mathbf{F}^{-T}\right] \colon \nabla\bm{v} \,dV - \int_{\partial_N\Omega_0} \Bar{\bm{T}} \cdot \bm{v} \,dA & = 0, \\
    -\int_{\Omega_0} \left(b(d)(J - 1) + \frac{p}{\kappa}\right) q \,dV & = 0, \\
    \int_{\Omega_0} \Bar{\lambda}_\mathrm{b} \beta \,dV 
    - \int_{\Omega_0} \left[\lambda_\mathrm{b}^{\text{max}} + \ll\lambda_\mathrm{b}^{\text{max}} - \lambda_\mathrm{b}\gg \right] \beta \,dV 
    + \int_{\Omega_0} g(d)\ell^2 \nabla\Bar{\lambda}_\mathrm{b} \cdot \nabla\beta \,dV & = 0.
\end{split}
\end{equation}

\section{Numerical implementation}\label{Section:NumericalImplementation}
The numerical implementation of the GED model is developed using the \texttt{FEniCS} finite element platform \cite{alnaes2015fenics}, leveraging the automatic differentiation features of the Unified Form Language (UFL). The corresponding Python scripts are publicly available on GitHub\footnote{\url{https://github.com/MMousavi98/Statistical_GED}}. To accommodate the near-incompressibility of the material response, a mixed finite element formulation is adopted. Following the methodology presented in \cite{li2020variational}, the displacement, pressure, and nonlocal Kuhn segment stretch fields are discretized as follows:

\begin{equation}\label{eq_displacement}
\bm{u}(\bm{x})= \sum_{i=1}^{N} N_i(\bm{x}) \bm{u}_i, \quad
p(\bm{x})= \sum_{i=1}^{N} N_i(\bm{x}) p_i, \quad
\Bar{\lambda}_\mathrm{b}(\bm{x})= \sum_{i=1}^{N} N_i(\bm{x}) \left(\Bar{\lambda}_\mathrm{b}\right)_i,
\end{equation}
where $N_i(\bm{x})$ are the shape functions associated with node $i$, and $\bm{u}_i$, $p_i$, and $\left(\Bar{\lambda}_\mathrm{b}\right)_i$ represent the nodal values of the displacement, pressure, and nonlocal Kuhn segment stretch, respectively. To ensure numerical stability and avoid spurious oscillations arising from the inf-sup condition in the coupled system of equations \eqref{weak_form_GED}$_1$ and \eqref{weak_form_GED}$_2$, a Taylor-Hood element pair is employed. This setup uses quadratic interpolation for the displacement field and linear interpolation for both pressure and nonlocal Kuhn segment stretch.

Assuming quasi-static loading, a staggered solution scheme is employed, advancing the simulation incrementally through a load-ramping function. This scheme consists of an outer iteration loop to monitor convergence, along with two inner loops to solve the staggered subproblems. In the first inner loop, the mechanical equilibrium and Lagrange multiplier equations (Eqs. \eqref{weak_form_GED}$_1$ and \eqref{weak_form_GED}$_2$) are solved simultaneously using the built-in \texttt{SNES} nonlinear solver in \texttt{FEniCS}, with the \texttt{newtontr} method. The nonlocal field $\Bar{\lambda}_b$ is held fixed at its value from the previous outer iteration, $\left(\Bar{\lambda}_\mathrm{b}\right)_{j-1}$
, where $j$ is the outer loop index. The solver uses a critical point (\texttt{cp}) line search with absolute, relative, and solution tolerances of $10^{-6}$, and a maximum of 300 iterations. In the second inner loop, the nonlocal Kuhn segment stretch equation \eqref{weak_form_GED}$_3$ is solved using the \texttt{vinewtonssls} method in \texttt{SNES}, while keeping the displacement and pressure fields fixed at $\mathbf{u}_j$ and $p_j$. The same solver tolerances and iteration limits are applied.

The outer loop checks convergence based on the supremum norm difference in the nonlocal stretch field, requiring $|\left(\Bar{\lambda}_\mathrm{b}\right)_j - \left(\Bar{\lambda}_\mathrm{b}\right)_{j-1}|_\infty < 2 \times 10^{-3}$. Upon convergence, the simulation proceeds to the next load step, using the current values of $\mathbf{u}$, $p$, and $\Bar{\lambda}$ as initial guesses. The outer loop is also capped at a maximum of $100$ iterations per load step.

The energy release rate is computed using a domain-based J-integral formulation \citep{li1985comparison}. This approach has previously been developed and applied by the authors in the context of large-deformation poroelasticity in elastomers \citep{bouklas2015effect}, and more recently for validating fracture energy predictions in phase-field models \citep{mousavi2024evaluating}. In the present study, we emphasize that, due to the nature of the damage formulation, the J-integral remains path-independent (analogous to the small-scale yielding assumption) provided that the damaged region does not intersect the integration domain on which the J-integral is calculated.
\section{Results and Discussion}\label{Section:results}
In the results section, we aim to first showcase the approach's robustness and consistency with methods that do not provide microscopic damage information. Additionally, we aim to understand the connection between the assumed nonlocal length scale of the material with the emerging fractocohesive length scale, and for the first time showcase their connection to the extent of microscopic damage, a task not as easily controlled and monitored in experiments. In the following simulations, the Helmholtz free energy function $\Psi$, the first Piola-Kirchhoff stress $\mathbf{P}$, and hydrostatic pressure $p$ are all normalized by shear modulus $\mu$. Additionally, all spatial dimensions and displacements are non-dimensionalized using the characteristic length of the domain $H$ in each problem.

\begin{figure}[h!]
    \centering
    \includegraphics[width=1\linewidth]{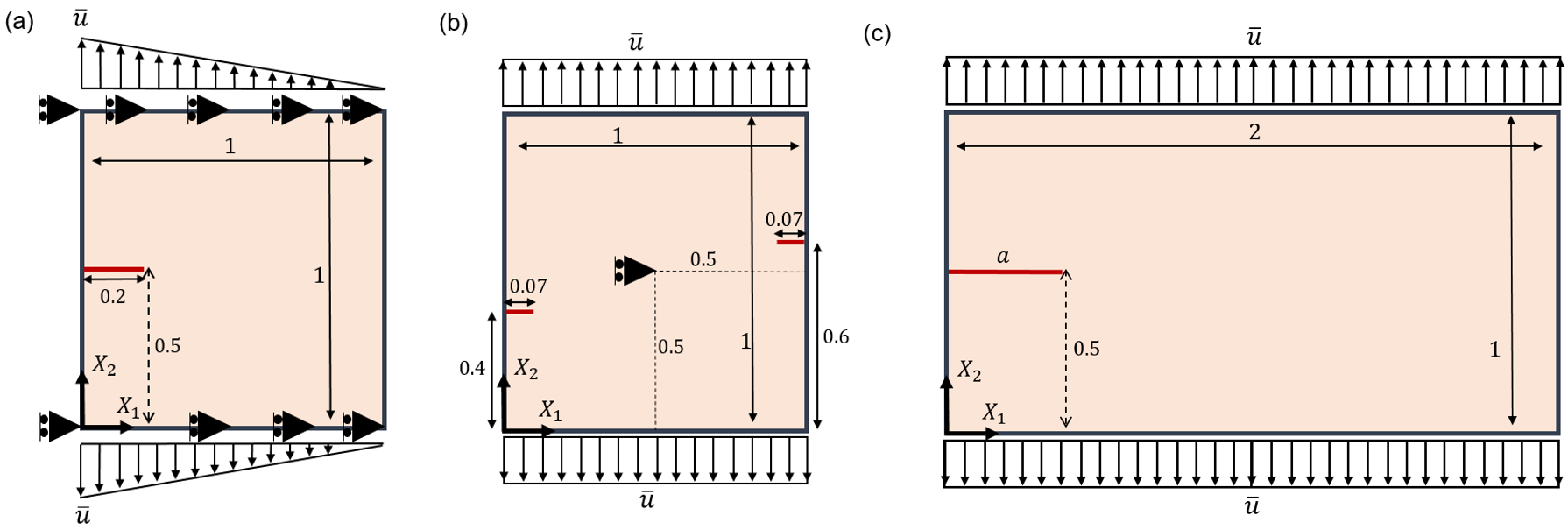}
    \caption{Square and rectangular domains under plane strain: (a) square domain with an edge crack under triangular displacement loading, (b) square domain with two discrete notches under uniform displacement, and (c) rectangular domain with an edge crack under uniform displacement.}
    \label{fig:loading}
\end{figure}

This study investigates three boundary value problems. In the first scenario, we consider a $L\times H=1 \times 1$ rectangular domain (with side length serving as the characteristic dimension), modeled under plane strain conditions and subjected to a triangular displacement loading, as illustrated in Fig. \ref{fig:loading}(a). A refined mesh with element size $h_e/\ell \approx 1/8$ is applied in the vicinity of the crack path \footnote{For details on a mesh sensitivity study using the proposed GED variant, the reader is referred to Mousavi \textit{et al.} (2025) \cite{mousavi2025chain}.}. A pre-existing diffuse notch of length $0.2$ is introduced at the center of the left boundary. The top and bottom boundaries are fixed in the $X_1$ direction. A prescribed displacement of $\Bar{u}_2 = 0.3(1 - X_1)$ is applied to the top boundary, while the bottom boundary is assigned $\Bar{u}_2 = -0.3(1 - X_1)$, with the origin located at the bottom-left corner of the domain. Displacements are incrementally applied over $300$ loading steps. The left and right boundaries are traction-free. For the nonlocal field $\Bar{\lambda}_\mathrm{b}$, a homogeneous Neumann boundary condition is enforced across all boundaries, i.e., $\nabla\Bar{\lambda}_\mathrm{b} \cdot \bm{n}_0 = 0$.

The second problem simulates crack propagation in a similar $1 \times 1$ square domain (with an unstructured mesh), also under plane strain conditions, but containing two discrete notches. The geometry and loading setup are depicted in Fig. \ref{fig:loading}(b). The notches are rectangular, with a width of $0.04$, a length of $0.07$, and rounded ends of radius $0.02$. Uniform displacements of $\Bar{u}_2 = 2$ and $\Bar{u}_2 = -2$ are applied to the top and bottom surfaces, respectively. Unlike the first case, these boundaries are free to move in the $X_1$ direction. To prevent rigid body motion, the displacement in the $X_1$ direction is fixed at the center point $(X_1, X_2) = (0.5, 0.5)$. As before, the lateral surfaces are traction-free, and $\nabla\Bar{\lambda}_\mathrm{b} \cdot \bm{n}_0 = 0$ on all boundaries. 

As illustrated in Fig. \ref{fig:loading}(c), the third boundary value problem involves a $2 \times 1$ rectangular domain ($h_e/\ell\approx0.04$) subjected to plane strain conditions. A pre-existing diffuse notch of length $a$ is introduced at the center of the left boundary. Uniform displacements of $\Bar{u}_2 = 3$ and $\Bar{u}_2 = -3$ are imposed on the top and bottom edges, respectively. All other boundary conditions are identical to those used in the second boundary value problem.

\begin{figure}[h!]
    \centering
    \includegraphics[width=0.6\linewidth]{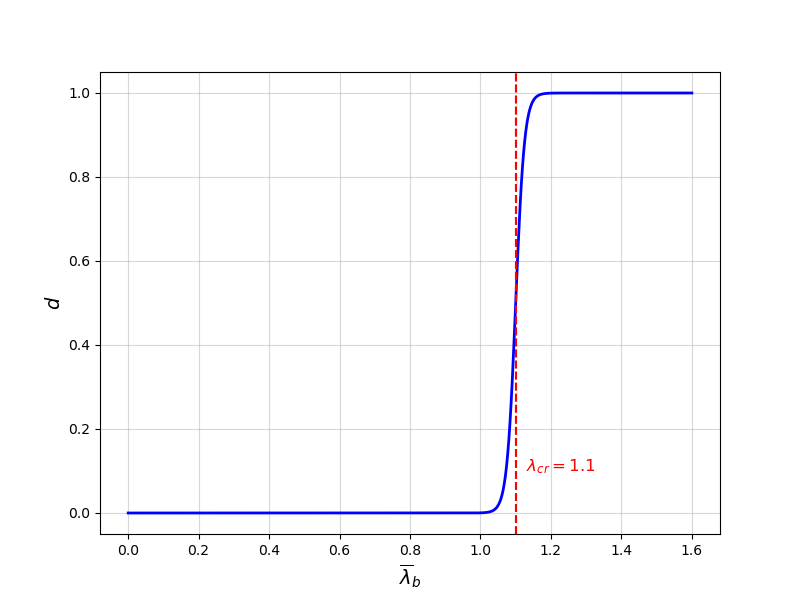}
    \caption{Damage function for $c = 80$ and $\lambda_\mathrm{cr} = 1.1$.}
    \label{fig:damage_function}
\end{figure}

Throughout all simulations, the bulk to shear modulus ratio $\kappa/\mu$ and bond stretch stiffness to shear modulus ratio $E/\mu$ are both set to $1000$. This corresponds to a Poisson’s ratio of $\nu = 0.499$ indicating near-incompressibility. The nondimensional nonlocal length scale is set to $\ell/H = 0.04$. Also, in all simulations except those specifically mentioned, $N=4$, $\lambda_{b}^{max}=1.5$, and $m=0.12$. The damage model is specified using a degradation parameter $c = 80$ and a critical stretch $\lambda_\mathrm{cr} = 1.1$, representing the midpoint of the damage transition. The resulting damage function is shown in Fig. \ref{fig:damage_function}. In the first and third boundary value problems, a pre-existing diffuse crack is introduced by assigning $\Bar{\lambda}_b = 1.2$ ($\Bar{\lambda}_b = 1$ everywhere else), which corresponds to a fully damaged state ($d = 1$) according to the damage law \footnote{Although the sigmoid formulation does not theoretically attain a value of 1, in practice the damage variable reaches 1 within machine precision in the numerical simulations.}. This condition also serves as the initial lower bound for the bounded \texttt{SNES} nonlinear solver (method: \texttt{vinewtonssls}) when solving Eq. \eqref{weak_form_GED}$_3$. In the second boundary value problem, damage is initiated through explicit geometric notches as described above, so the initial lower bound for the bounded \texttt{SNES} nonlinear solver is $\Bar{\lambda}_b = 1$ everywhere. A summary of the material and model parameters is provided in Table \ref{tab:parameters}.

\begin{table}[h!]
\caption{Material and model parameters used in the simulations.}
\centering
\renewcommand{\arraystretch}{1.2}
\begin{tabular}{|c|c|}
\hline
Parameter & Value \\ \hline
$\mu$              & 1              \\ \hline
$\kappa/\mu$       & 1000           \\ \hline
$E/\mu$            & 1000           \\ \hline
$N$                & 4              \\ \hline
$\ell/H$             & 0.04           \\ \hline
$\lambda_{b}^{max}$ & 1.5            \\ \hline
$m$                & 0.12           \\ \hline
$\lambda_\mathrm{cr}$     & 1.1            \\ \hline
$c$                & 80             \\ \hline
\end{tabular}
\label{tab:parameters}
\end{table}

The results section is organized into three parts. The first subsection focuses on simulating crack propagation for the first boundary value problem and extracting the corresponding fracture toughness (given the aforementioned material parameter settings for the model). This fracture toughness value is then used in the second boundary value problem to compare the results with predictions from the phase-field model. The second subsection investigates the physical interpretation of the internal length scale parameter $\ell$ and its role in flaw sensitivity. The third subsection presents an analytical estimation of the fracture toughness and compares it to the numerical results.

\subsection{Modeling crack propagation using GED and comparison with the phase-field method}\label{subsection:one}
\begin{figure}[h!]
    \centering
    \includegraphics[width=\linewidth]{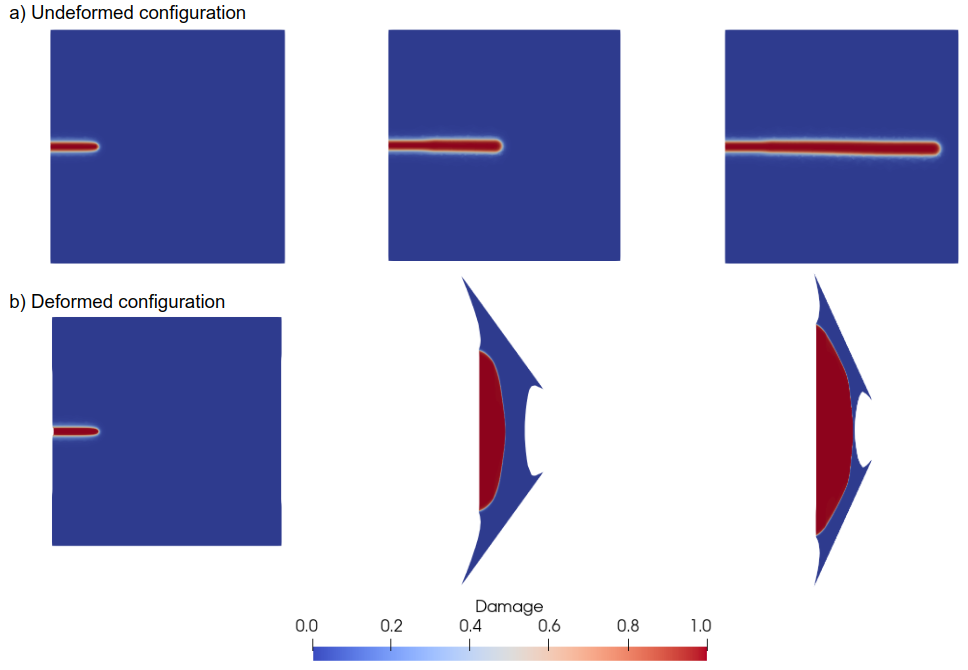}
    \caption{The result of the GED simulation for the first boundary value problem in the reference (top row) and current (bottom row) configurations at load steps 1, 137, and 220 (from left to right, respectively).}
    \label{fig:main_contours}
\end{figure}

\begin{figure}[h!]
    \centering
    \includegraphics[width=\linewidth]{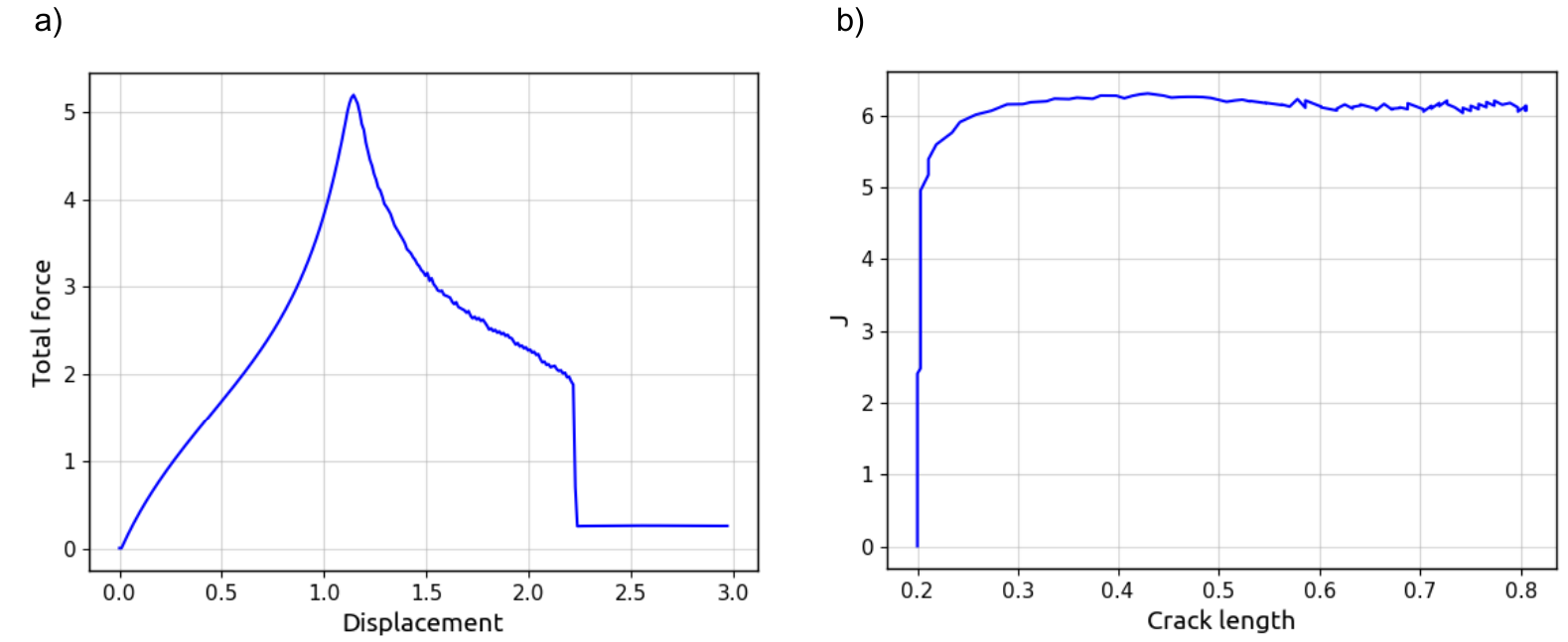}
    \caption{a) Force-displacement curve and b) energy release rate of the first boundary value problem}
    \label{fig:main_plots}
\end{figure}

In this section, we analyze the first boundary value problem depicted in Fig. \ref{fig:loading}(a), corresponding to Mode I loading. As previously mentioned, the initial damage is introduced in a diffuse manner. Simulation results at three representative load steps are shown in Fig. \ref{fig:main_contours}, illustrating both the undeformed and deformed configurations. As observed, no broadening of the damage zone is present. The corresponding force-displacement and J-integral curves are provided in Fig. \ref{fig:main_plots}. Based on the J-integral, the computed nondimensional critical energy release rate (fracture toughness) is approximately $G_c \approx 6.1$. It is important to note that this value is obtained as an outcome of the simulation, rather than being prescribed to the model in advance. This highlights a major advantage of the GED model over the phase-field method, which requires the fracture toughness to be specified \textit{a priori}.

\begin{figure}[h!]
    \centering
    \includegraphics[width=1\linewidth]{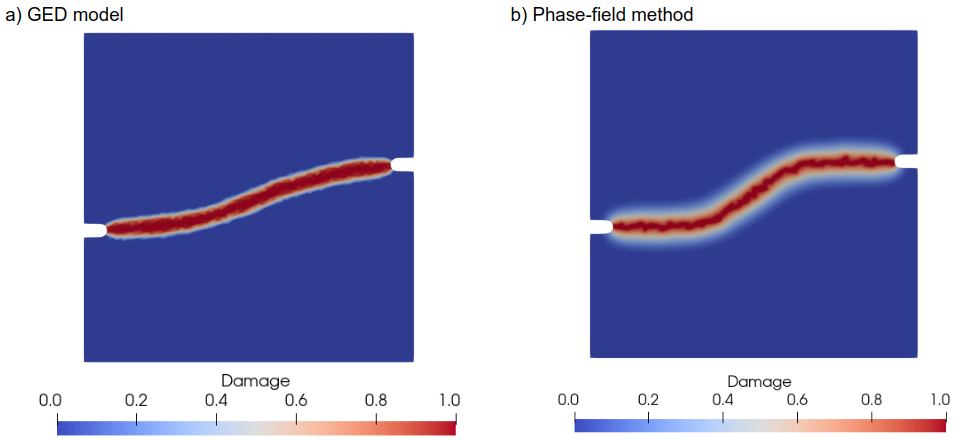}
    \caption{A comparison between the GED model and the phase-field method in modeling crack propagation for the second boundary value problem in the reference configuration.}
    \label{fig:two_notch_contours}
\end{figure}

\begin{figure}[h!]
    \centering
    \includegraphics[width=0.6\linewidth]{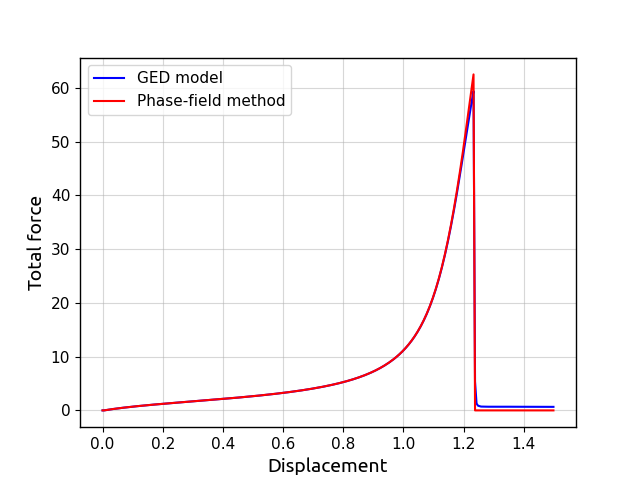}
    \caption{A comparison between the GED model and the phase-field method in modeling the crack propagation in the second boundary value problem based on the force-displacement curve.}
    \label{fig:two_notch_plots}
\end{figure}

Next, to evaluate the model in a different setting, we solve the second boundary value problem shown in Fig. \ref{fig:loading}(b). In this case, two discrete notches are introduced in the domain, and a uniform loading is applied, resulting in rapid crack propagation. The simulation results are compared against those obtained using a phase-field model. As mentioned above, the phase-field approach requires the fracture toughness to be specified in advance. For consistency, we use the value $G_c \approx 6.1$, determined from the first example via the J-integral, since all material parameters remain unchanged. The outcomes of both simulations are presented in Fig. \ref{fig:two_notch_contours}. In both models, crack propagation occurs in a single load step, and the resulting crack paths exhibit good agreement. To further validate the model, the force-displacement responses from both simulations are shown in Fig. \ref{fig:two_notch_plots}. The phase field approach does not capture a micromechanically meaningful representation of damage; it aims to attain a diffuse representation of the macroscopic fracture. Nevertheless, the match between the two curves provides additional confirmation of the predictive abilities of the GED model for the macroscopic response.\footnote{As previously discussed, the modified GED theory and model were thoroughly validated in Mousavi \textit{et al.} (2025) \citep{mousavi2024evaluating} through comparisons of energy dissipation curves, J-integral results, and phase-field simulations. Since the primary objective of the present work is to investigate the fundamental physics behind the model, the validation section is kept brief, with greater emphasis placed on the physical insight.
}

\subsection{Flaw sensitivity of elastomers and the role of the fractocohesive and nonlocal length scales}

\begin{figure}[h!]
    \centering
    \includegraphics[width=0.6\linewidth]{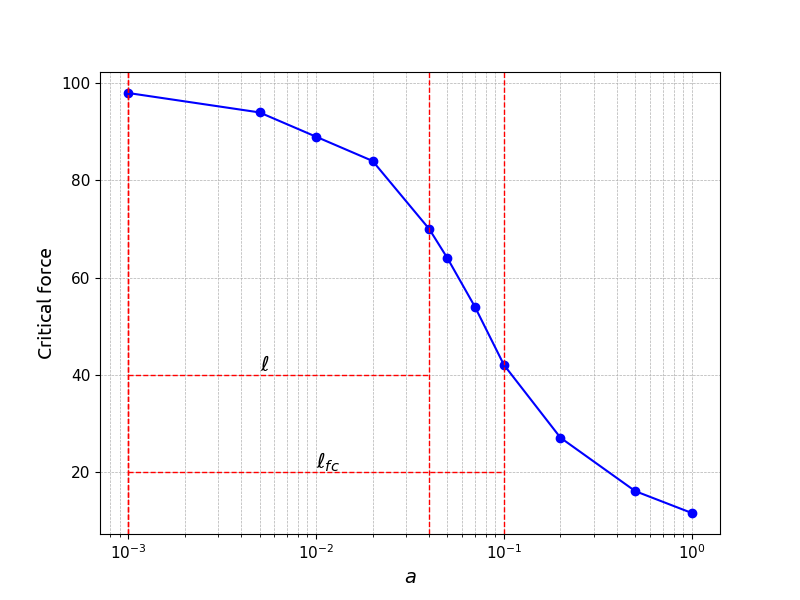}
    \caption{Critical force vs crack length curve to depict the flaw-insensitive region along with the two length scales.}
    \label{fig:flaw_sensitivity}
\end{figure}

Flaw sensitivity refers to the influence of a pre-existing defect on crack initiation and propagation. To investigate this effect, we consider the third boundary value problem illustrated in Fig. \ref{fig:loading} (c). A range of initial notch (flaw) sizes, varying from $a=10^{-3}$ to $a=1$, is introduced diffusely within the domain. For each case, the critical force required to initiate crack propagation is recorded, and the results are presented in Fig. \ref{fig:flaw_sensitivity}. Notably, when the notch length falls below the length scale $\ell$, the required force for crack initiation becomes largely independent of the notch size. This finding aligns with previous experimental and numerical observations reported in \cite{mao2017rupture, chen2017flaw}.

Previous studies \cite{mao2017rupture, chen2017flaw, long2021fracture} have suggested that the fractocohesive length (also known as dissipation length scale) is given by the ratio of fracture toughness $G_c$ to the work to rupture $W$, i.e., $\ell_{fc} = G_c / W$.\footnote{A physical intuition behind this formula is as follows: $W$ represents the energy needed to cause failure in a unit volume of material. To create a unit area of new crack surface, the material must dissipate energy over a process zone of thickness $\ell_{fc}$, giving an effective failure volume of $1 \times 1 \times \ell_{fc}$. The energy required to fracture this volume is then $\ell_{fc} W$, which must match the fracture energy $G_c$. Thus, $\ell_{fc} \sim G_c / W$.} The (nondimensional) fracture toughness, determined in the prior subsection, is $G_c \approx 6.1$. To evaluate the work to rupture, we set up a 3D variant of the second boundary value problem: a $1 \times 1 \times 0.04$ domain along the $X_1$, $X_2$, and $X_3$ directions (with an out-of-plane thickness of $0.04$), but without a pre-existing crack. The specimen was subjected to uniform uniaxial tension by loading at $X_2 = 0$ and $X_2 = 1$ until damage nucleation and rupture occurred, with the displacements in the $X_1$ and $X_3$ directions constrained to zero on these two facets. Note that all other $4$ faces were left unconstrained. In this case, the maximum elastic energy stored in the uncut sample (in the last recorded load step prior to the onset of damage \footnote{At a particular stage, the work to rupture attains its maximum value before decreasing due to damage initiation. It is this peak value that we report here.}) is measured to be $\int \Psi \, dV \approx 2.4$. Dividing this value by the domain volume ($1 \times 1 \times 0.04$) yields the work to rupture as $W \approx 60$. Consequently, the corresponding fractocohesive length scale is $\ell_{fc} \approx 0.1$.

\begin{figure}[h!]
    \centering
    \includegraphics[width=0.6\linewidth]{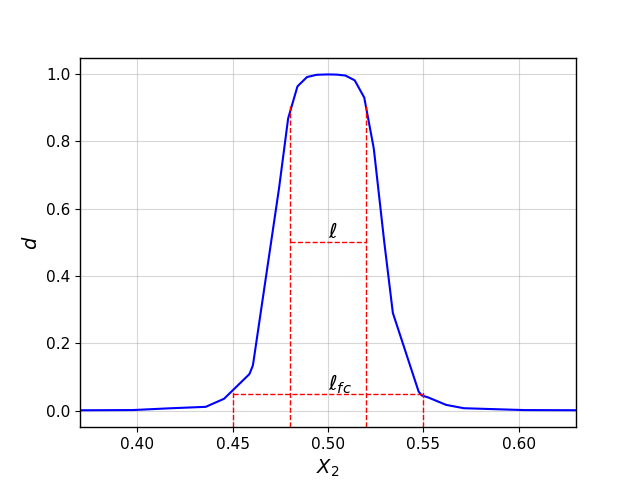}
    \caption{Damage profile at $X_1=0.5$ and $0.35\leq X_2\leq0.65$ after the full crack propagation in the first boundary value problem along with the two length scales visually inscribed (about the center of the crack).}
    \label{fig:damage_profile}
\end{figure}

Interestingly, this $\ell_{fc}$ value is approximately 2.5 times larger than the nonlocal length scale $\ell$ used in the simulations and in the flaw sensitivity analysis in Fig. \ref{fig:flaw_sensitivity}. It is important to clarify that while the nonlocal length scale $\ell$ influences the width of the damage zone, it is not (necessarily) equal to it. Fig. \ref{fig:damage_profile} compares the observed damage zone profile with the two identified length scales. Remarkably, each length scale corresponds to a different characteristic width within the damage zone: the smaller length scale $\ell$ aligns with the width of the highly damaged core region, whereas the fractocohesive length $\ell_{fc}$ corresponds approximately to the width of the entire damage zone (up to $d\approx 0.1$). This observation is consistent with the physical interpretation of $\ell_{fc} = G_c/W$, which represents the region where the irreversible dissipation process (i.e., bond scission) is occurring \cite{long2021fracture}.

\subsection{A connection to the Lake-Thomas model}

We now compare our approach with the well-known scaling laws for the critical energy release rate of elastomers, originally proposed by Lake and Thomas \cite{lake1967strength}. To this end, we utilize a modification to the Lake and Thomas model proposed by Mao \textit{et al.} (2017) \cite{mao2017rupture} that assumes a length scale $\ell_{m}=r_0$ (recall that $r_0=\sqrt{N}b$ is the unstretched chain length). According to this, fracture toughness can be written as \footnote{The scaling law arises from a simple, elegant reasoning: the energy per unit area required to propagate a crack by a length $\ell_m$ is given by the number of chains per unit area in front of the crack (scaling as $n\ell_m$) multiplied by the energy stored in a single chain at the point of failure.}:
\begin{equation}\label{Gc_mao0}
G_c^* \approx n\ell_m N\zeta_k
\end{equation}
where $\zeta_k$ represents the average energy required to rupture a segment. 
Interpreting the damage function defined in Eq. \eqref{damage_function} and shown in Fig. \ref{fig:damage_function} as a probabilistic representation of bond failure under a given stretch $\lambda_b$, we compute $\zeta_k$ by considering the bond energy evolution as damage is accumulated up to the point of scission:
\begin{equation}\label{zeta}
\zeta_k = \int_{0}^{1} \frac{1}{2} E_b \, (\lambda_b - 1)^2 \, dd.
\end{equation}
Note that this calculation of $\zeta_k$ is in line with the physical picture of force-coupled polymer chain scission described by Wang et al. (2019) \cite{wang2019quantitative} and highlighted within the recent work of Wang et al. (2025) \cite{wang2025loop}. Further, recall that $E_b = E/n$, where --working with nondimensional quantities-- $n=2.41 \times 10^{26}$ is the chain density. Since  $E = 1000$, then $E_b = 1000 / (2.41 \times 10^{26}) = 4.15 \times 10^{-24}$. Substituting this into Eq. \eqref{zeta} and performing the integration yields $\zeta_k = 2.18 \times 10^{-26}$ (or $0.53\%$ of $E_b$). Also, to compare with the numerical prediction, we set $\ell_{m}=\ell=0.04$ and recall that $N=4$. Finally, using Eq. \eqref{Gc_mao0}, we estimate the nondimensional fracture toughness as $G_c^* \approx 0.842$. If we instead take $\ell_m$ in Eq. \eqref{Gc_mao0} to be equal to the fractocohesive length of $\ell_{fc} = 0.1$, then the estimate improves to $G_c^* \approx 2.1$, which is closer to the numerical result of $G_c \approx 6.1$. Although a discrepancy remains between the analytical and numerical values, it is important to recognize that the analytical expression serves only as an approximation and is not expected to yield exact results. Even so, the fact that the estimates are within the same order of magnitude as the numerical result is encouraging, suggesting that the analytical formula can offer useful insight, particularly in experimental contexts. In this vein, if $G_c$ and $W$ are obtained experimentally, one can deduce the fractocohesive length $\ell_{fc}$ as previously described (via $\ell_{fc} = G_c / W$). Subsequently, Eq. \eqref{Gc_mao0} can be used (with $\ell_{m} = \ell_{fc}$) to determine $N \zeta_k$, representing the average energy dissipated per chain. Then, given the molecular weight of the chains in the network, an estimate of the number of Kuhn segments per chain can be obtained, allowing the energy dissipation per segment to be further estimated by dividing $N \zeta_k$ by the number of segments.

\begin{figure}[h!]
    \centering
    \includegraphics[width=\linewidth]{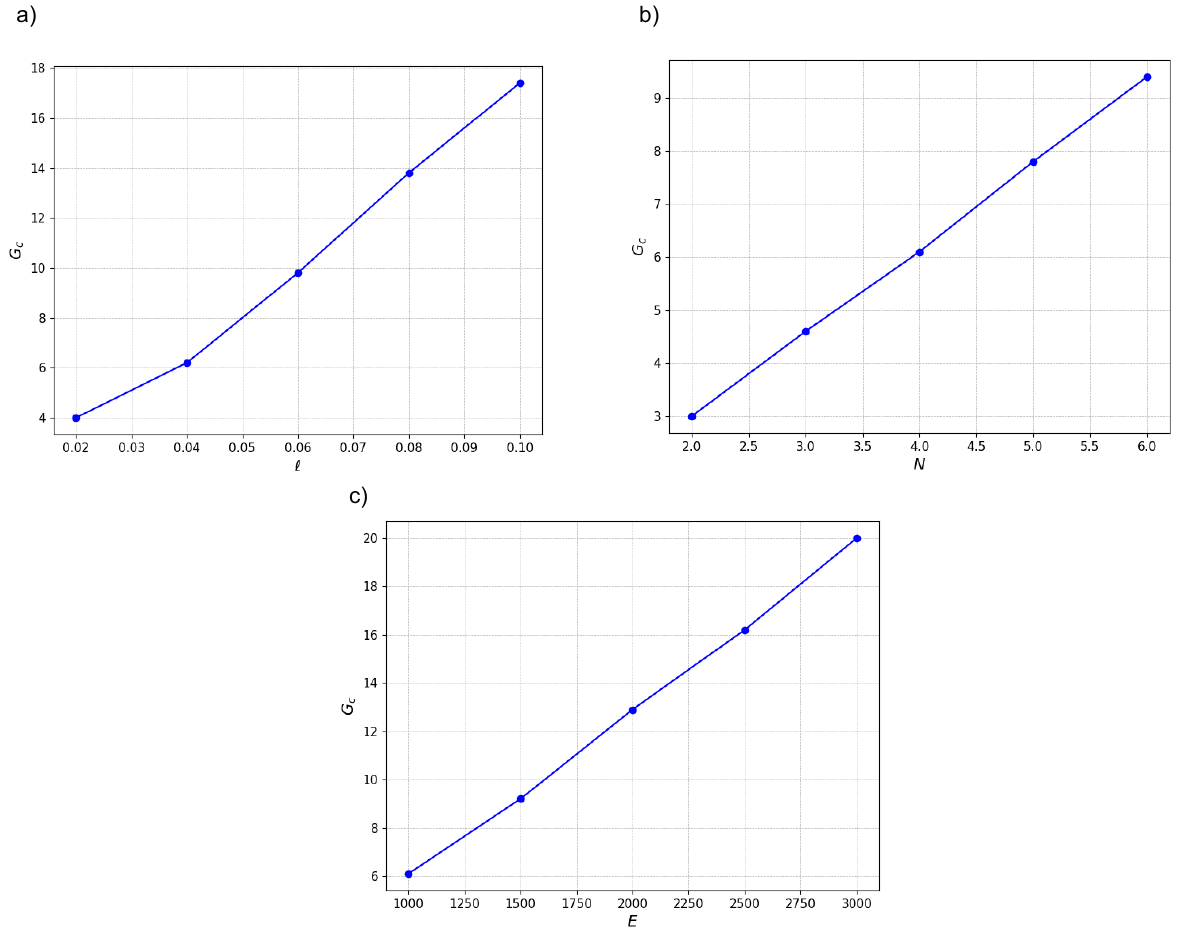}
    \caption{The effect of having different a) length scale $\ell$ while $N=4$ and $E=1000$, b) the number of Kuhn segments $N$ while $\ell=0.04$ and $E=1000$, and c) the network stiffness $E$ while $\ell=0.04$ and $N=4$ on the fracture toughness $G_c$.}
    \label{fig:param_study}
\end{figure}

Furthermore, we utilize our numerical model to perform a parametric analysis of the fracture toughness $G_c$ with respect to the length scale $\ell$, the number of Kuhn segments $N$, and the network stiffness $E$. As shown in Fig.~\ref{fig:param_study}, the results reveal a linear relationship between $G_c$ and each of these parameters. Interestingly, these trends are consistent with the linear relationship of Eq. \eqref{Gc_mao0}, reinforcing the validity of the analytical formulation.



\newpage
\section{Conclusion}\label{Section:conclusion}
In this work, we incorporated the Helmholtz free energy density derived from a monodisperse elastomer network into a stretch-based gradient-enhanced damage (GED) framework to simulate crack propagation in near-incompressible elastomers. This approach allowed us to embed statistical mechanics insights of polymer chains directly into the continuum model. 
To ensure thermodynamic consistency, the governing equations were derived using the principle of virtual power, and the constitutive relations were formulated in accordance with the Clausius–Duhem inequality. The resulting model successfully captured crack initiation and propagation without spurious broadening of the damage zone. Moreover, the fracture toughness emerged as an output of the simulation, while the physically meaningful damage distribution of polymer chains was captured spatially, enabling a direct comparison with a corresponding phase-field simulation (using the emergent fracture toughness value from our stretch-based GED model). The study also confirmed that when the flaw size is smaller than the characteristic length scale of the elastomer -- which corresponds to the nonlocal length scale of our GED model -- the force required to initiate a crack becomes relatively insensitive to the flaw size, a behavior consistent with the known flaw-insensitive response of elastomers. Finally, by dividing the fracture toughness by the work to rupture, we determined the fractocohesive length of the material. Our model highlights the connection of the nonlocal length scale (simulation input) and the emerging fractocohesive length scale (simulation output) to the network level chain damage distribution. Specifically, while the nonlocal length scale pointed to a width where the percolation threshold has been exceeded, the fractocohesive length scale pointed to a broader domain that coincided with a region where chain scission takes place, as predicted by the numerical simulations. This is the first time a model can provide insight into the mechanical underpinnings of chain scission, damage, fracture, nonlocal modeling features due to network-level heterogeneity, and the emergence of the fractocohesive length scale.

\section{Acknowledgments}
SMM and NB acknowledge the support by the National Science Foundation under grant no. CMMI-2038057. JPM gratefully acknowledges the support of the National Science Foundation Graduate Research Fellowship Program under Grant No. DGE-1650441. Any opinions, findings, conclusions, or recommendations expressed in this material are those of the author(s) and do not necessarily reflect the views of the National Science Foundation. JPM also gratefully acknowledges the support of the Air Force Research Laboratory Materials and Manufacturing Directorate, and the National Research Council (NRC) Research Associateship Program (administered by the National Academies of Sciences, Engineering, and Medicine). The work of BT was performed under the auspices of the U.S. Department of Energy by Lawrence Livermore National Laboratory under Contract DE-AC52-07NA27344.

\section{Highlights}\label{Section:highlights}
\begin{itemize}
  \item A thermodynamically consistent gradient-enhanced damage model is formulated using the Helmholtz free energy of a monodisperse polymer network.
  \item A smooth sigmoid-based damage law is introduced to ensure robust numerical performance.
  \item The model captures flaw-insensitive fracture behavior when notch sizes are smaller than the imposed length scale parameter.
  \item The full width of the damage zone is, for the first time, shown to coincide with the fractocohesive length, derived from the ratio of fracture toughness to work to rupture.
\end{itemize}


\end{document}